# Scalable heterostructures produced through mechanical abrasion of van der Waals powders

*Darren Nutting[1], Jorlandio F. Felix[2], Dong-Wook Shin[3], Adolfo De Sanctis[1], Monica F Craciun[1], Saverio Russo[1], Hong Chang[1], Nick Cole[1], Adam Woodgate[1], Ioannis Leontis[1], Henry A Fernández[1] and Freddie Withers[1\**

[1] *Centre for Graphene Science, College of Engineering, Mathematics and Physical Sciences, University of Exeter, Exeter EX4 4QF, United Kingdom*

[2]*Universidade de Brasília – UNB, Instituto de Física, Núcleo de Física Aplicada, 70910-900 Brasília, DF, Brazil*

[3]*Electrical Engineering Division, Department of Engineering, University of Cambridge, 9 JJ Thomson Avenue, Cambridge, CB3 0FA, UK*

E-mail: f.withers2@exeter.ac.uk

**Abstract**

To fully exploit van der Waals materials and heterostructures, new mass-scalable production routes that are low cost but preserve the high electronic and optical quality of the single crystals are required. Here, we demonstrate an approach to realize a variety of functional heterostructures based on van der Waals nanocrystal films produced through the mechanical abrasion of bulk powders. Significant performance improvements are realized in our devices compared to those fabricated through ink-jet printing of nanocrystal dispersions. To highlight the simplicity and scalability of the technology a multitude of different functional heterostructure devices such as resistors, capacitors, photovoltaics as well as energy devices such as large-area catalyst coatings for hydrogen evolution reaction and multilayer heterostructures for triboelectric nanogenerators are shown. The simplicity of the device fabrication, scalability, and compatibility with flexible substrates makes this a promising technological route for up-scalable van der Waals heterostructures.

# 1. Introduction

High quality van der Waals (vdW) heterostructures offer many improvements over conventional materials. They are lightweight, semi-transparent and are compatible with flexible substrates whilst at the same time display competitive performance to that of conventional compound semiconductor materials. The highest quality heterostructures are mainly still created through mechanical exfoliation of bulk single crystals and built up layer-by-layer by standard mechanical transfer procedures.[1-3] This method, however, is not scalable and alternative device manufacturing routes are sought after for industrial applications.

For instance, chemical vapour deposition (CVD)[4], where monolayer films are grown layer by layer at high temperatures have shown promising results with similar performance to exfoliated crystals. However, the energy cost of growth is high for a given quantity of material, the growth of multi-layer heterostructures is complex and is also confined to a small number of material combinations. Furthermore, transfer of the films from catalyst substrates often introduces contamination, tears and cracks which prevents the formation of high quality vertical heterostructure devices.

An alternative low-cost route for the mass-scalable production of nanocrystal heterostructures is through the printing of liquid-phase exfoliated dispersions.[5-7] In this scheme, the 2D dispersions are produced through either ultra-sonication or shear force exfoliation of bulk van der Waals microcrystals in suitable solvents[8]. This leads to stable dispersions which can then subsequently be printed on a variety of substrates. By mixing the dispersions with specialist binders the heterostructures can also be built-up layer-by-layer through ink-jet printing[5]. However, strong disorder in the crystals caused by oxidation, small crystallite size and poor interface quality leads to severe performance degradation compared to devices based on mechanically exfoliated or CVD-grown crystals. Furthermore, this production method is unlikely to be compatible with many highly air sensitive vdW materials, limiting the scope of the technology. Moreover, residue solvent in the printed films has been shown to degrade the electrical properties of the devices by further reducing the quality of the interface between neighbouring nanocrystals[9].

This works sets out a route to build up semi-transparent and flexible vdW nanocrystal heterostructures through a scalable, solvent-free technique which is suitable for all layered materials with no restriction on their air-stability. This technique is based on mechanical abrasion process. Here, we show that this process can be used to create various electronic and optoelectronic heterostructures that can be readily fabricated within a matter of minutes on the scale of 10's of cm scale and could easily be scaled up further.  Specifically, in this work we focus on several van der Waals materials including graphite, $MoS_2$, $WS_2$, $MoSe_2$ and hexagonal boron nitride (hBN).  We show several examples of functional electronic and optoelectronic heterostructures including multi-layer graphene field effect transistors, vertical transition-metal dichalcogenide (TMDC) photodetectors, photovoltaics, hBN capacitors, hydrogen evolution catalysts (HER) and multilayer films for triboelectric nanogeneration (TENG).

## 2. Results
### 2.1 Device characterisation and fabrication

The general approach used to produce thin films and devices based on abraded 2D nanocrystals on $SiO_2$ as well as polymer substrates is shown in Figure 1A.   We make use of a soft polymer, namely Polydimethylsiloxane (PDMS) which is coated in a 2D material powder film and used as a writing pad for the nanocrystal films. The PDMS pad is oscillated back and forth against the substrate with various 2D materials embedded between it and the substrate. To ensure that the 2D material is only written at selective locations, a tape mask is applied to the substrate before writing. After the material has been written the tape is removed leaving only the unmasked region coated in the 2D material, Figure 1A (Step 1 to 3). This process can then be repeated to build up heterostructures, Figure 1A (Step 4). An example of a set of connected vertical heterostructures produced through fabrication route 1 is shown in the top micrograph of Figure 1C and in the supplementary information. Following route 1, we found that vertical structures with a top graphitic electrode suffered from short-circuits most probably due to intermixing of the top graphitic layer with the barrier material. To overcome this, we have also developed a separate route, which involves transferring the graphitic top electrode directly onto the barrier material. This is achieved by abrading graphite directly onto a PMGI polymer layer followed by a spin coated PMMA layer. The sacrificial PMGI is then subsequently dissolved in a bath of MF319 leaving the graphitic film attached to the Poly-methyl-methacrylate (PMMA) layer which can then be transferred onto the target substrate. This fabrication route is schematically illustrated in Figure 1B.

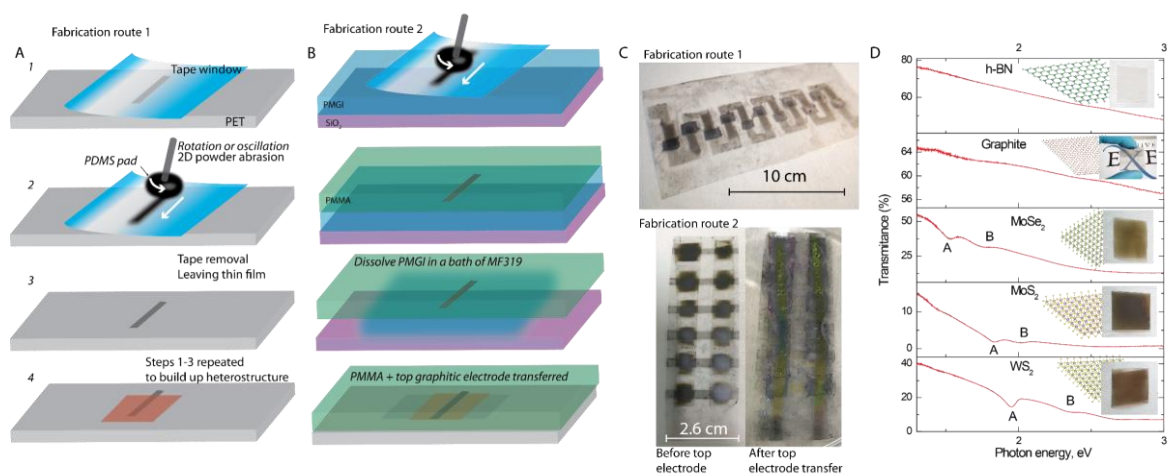

**Figure 1. Thin film produced through powder abrasion.** (A) Fabrication route to produce heterostructures through mechanical abrasion of vdW powders via a direct write method. (B) Alternative fabrication route with transferred nano-crystal films. (C) (top) An example of multi-layered vertical junction photodetectors based on graphite-WS2-graphite. (bottom) The same architecture as the top micrograph by this time following fabrication route two, with the top graphitic electrode transferred, which leads to a higher device yield. (D) Optical transmission spectra for 2D material coatings on PET substrates.

After device fabrication we characterise our films through a combination of optical and Raman spectroscopy[10-13], electron transport, atomic force and scanning electron beam microscopy to identify the crystallite sizes, roughness and film thicknesses (See supplementary information). Figure 1D shows the optical transmission spectra of several different abraded 2D material coatings on a 0.5 mm thick Polyethylene terephthalate (PET) substrate, the resultant films still display some level of transparency which can also be tuned to higher levels by back peeling the films with an adhesive tape. Examples of the transmission spectra of various films are also shown in Figure 1D. For the transition metal dichalcogenide films, strong absorption associated with the A and B excitons are still observed, indicating that the films still display similar optical properties comparable to the bulk pristine exfoliated crystals.[14, 15]

Important for applications are the electronic performance of the films. Figure 2 highlights some basic electrical characterisation measurements of the films (all measurements are performed in ambient conditions). Figure 2A displays an I-$V_b$ curve for a thin graphite film with transparency of 20% which gives a characteristic resistance per square of 2 KOhm/sq, indicating crystallite

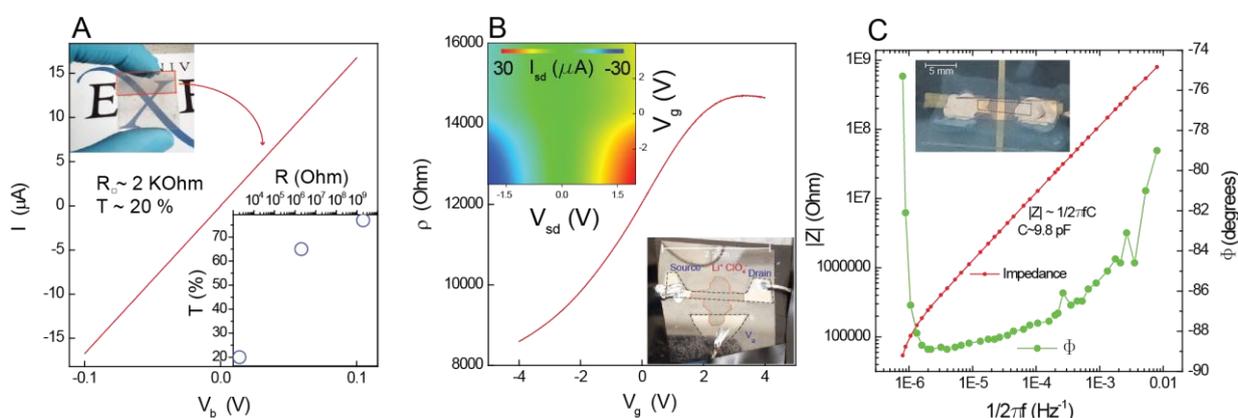

**Figure 2. Electronic properties of the bare films.** (A) I-$V_b$ of a graphitic film produced through mechanical abrasion (inset top left: optical photograph of a graphitic film produced on a 2.5 cm PET substrate thinned to three different thicknesses by back-peeling with adhesive tape. Inset bottom right: Transparency vs resistance for different thickness graphitic films) (B) Gate dependence of the channel resistivity for the device shown in the bottom-right inset using an liClO$_3$ electrolyte (scale bar = 2.5 cm). Top left inset: Contour map of the $I_{sd}$-$V_{sd}$ for different applied gate voltages. (C) Impedance spectroscopy for a hBN dielectric capacitor produced using a 5μm thick hBN film.

thicknesses of ~10 L.

We find the value of resistance to be similar to previous reports who focus on planar graphitic films which are produced through a more complex multi-step production process[16]. Importantly, the thickness of the film can be controlled by thinning with specialist tapes (see Methods), allowing for improved transparency but increased resistance. In thicker abraded graphitic films, we can ultimately obtain resistances down to 100 Ohm/sq. This could ultimately be lowered further to the 10's Ohm/sq through various doping methods such as intercalation[17], making such films useful for electrodes in optoelectronic devices.

Interestingly, it is found that the resistance of the few-layer graphitic channel can be controlled by application of a gate-voltage, in this case we employ an electrolyte gate, lithium perchlorate (Li$^+$:ClO$_3^-$)[18], which has been drop cast over the channel region and contacted using an abraded graphitic gate electrode. Figure 2B shows the typical resistance vs gate voltage for a tape thinned graphitic channel region, with the inset showing a contour map of the I$_{sd}$-V$_{sd}$ for different applied gate voltages. We find that the electro-neutrality region is at large positive gate voltages indicating a strong p-type doping[19], likely due to ambient water or oxygen doping. We found that all the studied abraded 2D materials are compatible with a wide variety of substrates such as, PET, PMGI, PTFE, PDMS, SiO$_2$ and paper. With the exception of graphite which was found to be incompatible with SiO$_2$ substrates.

Dielectric barrier layers are key elements of many electronic devices, ranging from field-effect transistors to light-emitting devices. Here, we make use of hexagonal boron nitride (hBN) dielectrics produced through mechanical abrasion over gold (Au) electrodes which were previously deposited by thermal evaporation, this results in dielectric films of thickness $5 \pm 1\ \mu m$ (estimated from AFM and surface profile measurements). Following the deposition of the hBN dielectric, a sheet of chemical vapour deposited (CVD) graphene is transferred onto the hBN film (see Methods) with two Au electrodes which act as the source and drain contacts for the graphene channel, the optical image of the assembled capacitor is shown in the inset of Figure 2C. The total area of the capacitor in this instance was estimated to be $2 \times 10^{-6}\ m^2$. The impedance spectrum is presented in Figure 2C and can be well described by the capacitive contribution, $|Z_T| = (2\pi f C)^{-1}$ at low frequency. The gradient to the linear fit gives the capacitance, which we find to be $C = 9.8\ pF$. Assuming a parallel-plate model, in which the capacitance is related to the relative dielectric permittivity, $\varepsilon_r$ via $C = \frac{\varepsilon_r \varepsilon_0 A}{d}$, we can estimate $\varepsilon_r \sim 2.24 - 3.36$. Previous reports have found widely varying values of the dielectric constants for nanocrystal hBN dielectrics with values ranging from 1.5 up to 200[20-23], whilst single crystal hBN is known to possess values ∼4.[24] The lower value of $\varepsilon_r$ in the abraded material could be due to air voids which will lower the effective dielectric constant of the mixed dielectric film.

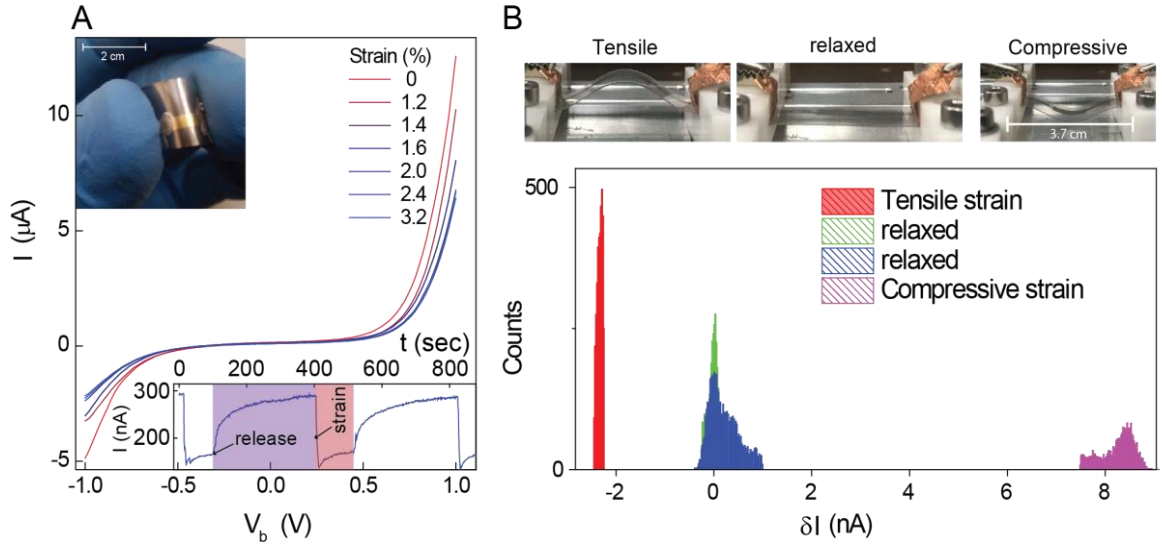

**Figure 3. Strain dependent properties of abraded films.** (A) Typical I-$V_b$ curve for a 5 mm x 0.025 mm two-terminal planer $WS_2$ abraded film with an average film thickness of 1 $\mu$m for different levels of applied uniaxial strain. Bottom inset: time-evolution of the current though the device during a strain cycle ($V_b$ = 0.5V). (B) Compressive and tensile strain dependent changes in the current through a vertical graphite-$MoS_2$-WS2-graphite junction for over $10^3$ bending cycles at a strain level of 4%.

We also performed similar electrical characterisation of vdW heterostructures under strain, Figure 3A shows some typical I-$V_b$ curves for a planar graphite-$WS_2$-graphite channel on a PET substrate. The different curves are for increasing (red to blue) uniaxial strain generated by bending the 0.5 mm PET substrate in a custom-built bending rig (See supplementary information). We find that the device resistance increases for increasing levels of tensile strain and drops for increasing compressive strain, expected as the nanocrystals are being separated and compressed together, respectively. We also find that the resistance changes are highly reversible, see lower inset of Figure 3A, indicating the suitability of these films for strain sensing applications.

To determine the long-term durability of such strain sensors, we performed over $10^3$ bending cycles for both tensile and compressive strain for a fixed bias voltage. Figure 3B shows histograms for the change in current through the channel for both compressive and tensile strains of ~4%.

## 2.2 Photodetection and photovoltaic devices

TMDC's are semiconductor materials with an indirect bandgap in the bulk which have already shown great promise for future flexible photovoltaic and photodetection applications.[25-28] Heterostructures based on liquid phase exfoliated nanocrystals typically display poor photo detectivity in the order of 10-1000 $\mu$A/W restricting their use in practical applications.[5, 21, 29-32]

Here we fabricated multiple photodetector devices by mean of mechanical abrasion of TMDC crystals, as shown in Figure 4. A multitude of vertical and planar devices has been studied. Starting with the most basic, we focus on simple in-plane $WS_2$ channels on Si-$SiO_2$ substrates.

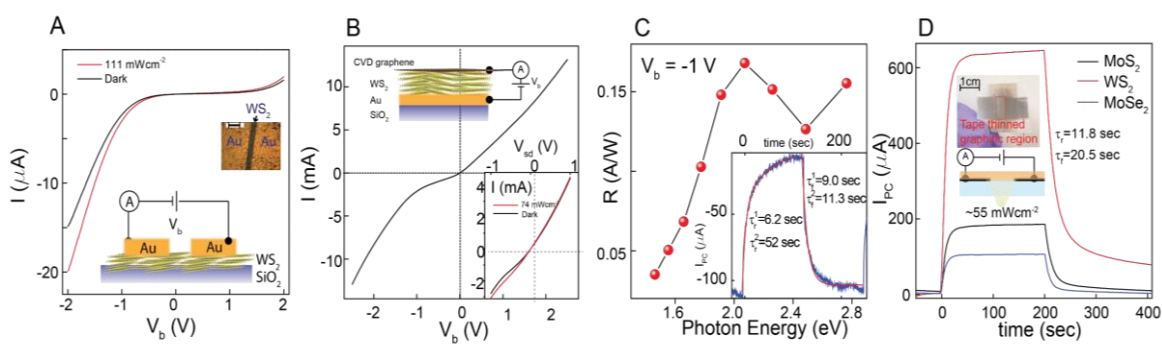

**Figure 4. Mechanically abraded films for photodetection applications.** (A) I-$V_b$ for a planer geometry $WS_2$ channel without (black curve) and with (red curve) 111 mW/cm$^2$ incoherent optical excitation. (B) I-$V_b$ for an Au-$WS_2$-CVD graphene top electrode with device area of 1 mm x 1 mm and $WS_2$ film thickness of ~1 $\mu$m. The inset shows a zoomed in region around $V_b$=+/-1V with and without optical excitation and with power density of 74 mW/cm$^2$. (C) Spectral dependence of the responsivity for the device shown in (B). (D) Temporal response for three planer photodetectors abraded onto PET substrates, consisting of Graphite-$MoS_2$, Graphite-$MoSe_2$ and Graphite-$WS_2$.

These devices were fabricated by mechanically abrading $WS_2$ nanocrystals over a 1 cm substrate followed by shadow mask evaporation of Cr/Au electrodes. Leaving a 2.5 mm channel length with a channel width of 1 mm and a typical film thickness of ~1 $\mu$m. The I-$V_b$ curves are shown in Figure 4A with (red curve) and without (black curve) excitation with white light of power density of 111 mW/cm$^2$. Such simple planar photodetectors already show improved photo-responsivity compared with liquid phase exfoliated materials with values up to 1.8 mA/W at $V_b$=-2 V. To better investigate the optoelectronic properties of our $WS_2$ abraded films, we employ CVD-grown graphene electrodes given their higher conductivity and optical transparency compared to abraded graphitic electrodes. The I-$V_b$ curve of a typical device is shown in Figure 4B and shows a much larger conductivity compared to the planar device in 4A. Indeed, similar vertical devices based on n- and p-type silicon show similar conductance with CVD-grown graphene top electrodes (See supporting information). The asymmetry in the I-$V_b$ curves here is due to the difference in the work functions of the graphene (4.6-4.9 eV) and Au (~5.2 eV) with the conduction band edge of the $WS_2$ closely aligned with the neutrality point of graphene.[33] This means that the conductivity is high at zero bias, as electron transport occurs through the conduction band of the $WS_2$[34]. At negative voltages, the energy difference between the chemical potential of graphene and the conduction band of $WS_2$, results in an increased barrier and lower conductivity. Figure 4C shows the spectral dependence of the photoresponsivity for this device with a peak responsivity at 2.0 eV consistent with the peak in absorption associated with the A-exciton, and maximal responsivities of 0.15 A/W at $V_b$=-1 V, this constitutes a $10^3$ enhancement over liquid-phase based photodetectors[5,20,28-31]. The time response to the incident white light source is also shown in the inset with peak photocurrent values of 100 $\mu$A at $V_{sd}$ = -1 V under uniform white light illumination of 74 mW/cm$^2$. An alternative photodetection architecture is shown in Figure 4D. This device consists of a tape-thinned graphitic channel directly coated with a TMDC layer. Similar bilayer devices have been reported previously and they typically consist of graphene/semiconductor[35-37] or graphene hybrids such as graphene/PbS quantum dot stacks.[38] Here we realise planar photodetectors by producing graphite-TMDC bilayers, where the TMDC is either $MoSe_2$, $MoS_2$ or $WS_2$. Figure 4D shows the temporal response for different TMDC layers. In order to ensure high enough optical transparency a thin strip of the graphitic channel was thinned with tape before application of the TMDC layer. The optical excitation was then carried out through the transparent PET substrate, we obtain a maximum responsivity of 24 mA/W for such scalable and simple detectors based on $WS_2$.

We now move our attention to more complex vertical heterostructure devices. Firstly we focus on graphite-$WS_2$-$MoS_2$-graphite heterostructure diodes with the top graphite electrode mechanically transferred using fabrication route 2 (see Figure 1B). An optical micrograph of a simple photodetector device is shown in Figure 5A, where different layers are highlighted for clarity. Figure 5B shows the I-$V_b$ curves for three such diode structures (D1-D3). The inset shows the Ohmic top and bottom graphitic electrodes I-$V_b$ characteristics with typical resistances of a few KOhm.

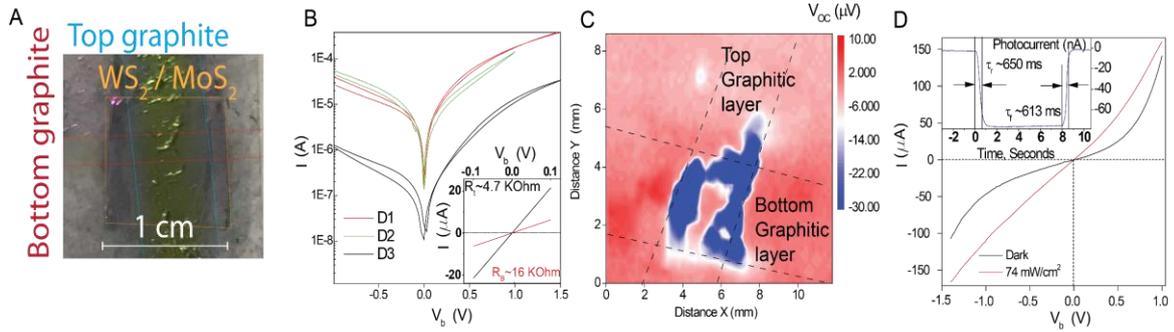

**Figure 5. Optoelectronics of abraded van der Waals diodes.** (A) Optical micrograph of a typical device. (B) I-$V_b$ curves for three representative devices. Inset: I-$V_b$ curves for the top and bottom graphitic electrodes. (C) Photo-voltage map of one of our diode structures measured with a focussed E=3.05 eV laser, power 0.5 mW and spot size of 5 mm. (D) I-$V_b$ curves for the device D2 shown in (B) with (red curve) and without (black curve) white light excitation of 74 mW/cm$^2$. Inset: temporal response of the short circuit photocurrent at $V_b$=0 V.

To indentify the photo-active region in the stack we have performed photovoltage mapping measurements on several devices as shown in Figure 5C and in the supplementary information, we always see peak values of the photovoltage on the overlap of all three materials indicating vertical electron transport as the dominant mechanism, the inhomogeneity arises due to variation of the contact quality of the top graphitic electrode. We note that the uniformity can be improved through annealing (See supplementary information).

Figure 5D shows the I-$V_b$ curves in dark (black) and under white light illumination. In these devices we measure responsivities 4-10 mA/W at $V_b$=-1.0 V and a zero bias photocurrent with response times of ~ 650 ms, significantly faster than the response times found in our planar devices, expected due to a much reduced electrode seperation.

## 2.3 Hydrogen evolution reaction (HER)

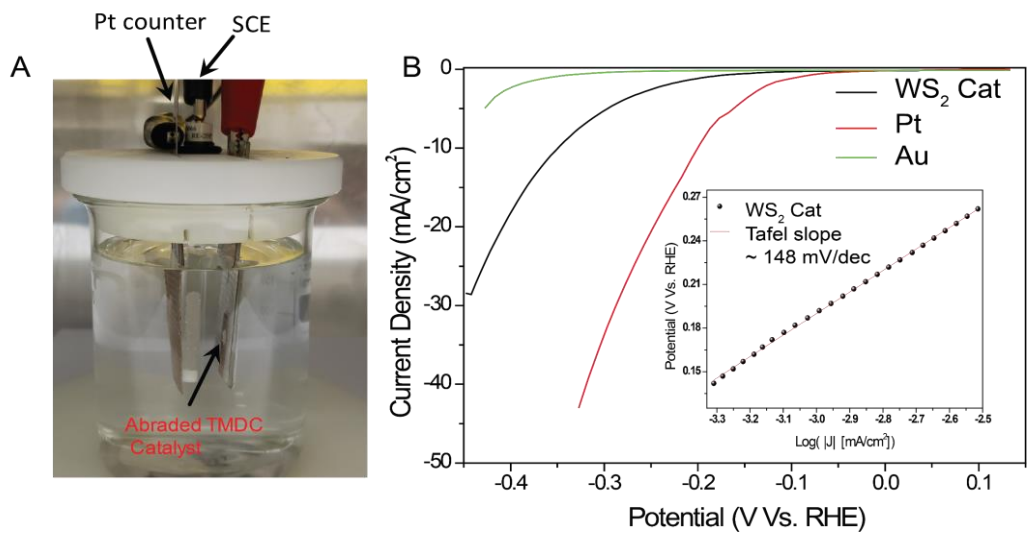

**Figure 6. WS$_2$ films as a catalyst for hydrogen evolution and photocurrent generation.** (A)Optical micrograph of the electrochemical cell highlighting the different electrodes (B) Polarization curves comparing Pt, Au and abraded WS$_2$ measured in 0.5 M H$_2$SO$_4$ with a scan rate of 2 mV/s at room temperature. The inset shows the Tafel plots for WS$_2$ sample.

Mono- and few- layer TMDCs have been widely studied for their potential as electrocatalysts for the hydrogen evolution reaction.

The electrochemical performance of our WS$_2$ films have been characterised in 0.5 M H$_2$SO$_4$ via linear sweep voltammetry (LSV).[39] To assess the suitability of abraded layered films as catalysts in HER, we have utilized, a three-electrode electrochemical cell was utilized where a Teflon tape was used to delimit the catalyst area (Figure 6A). For comparison, a commercial platinum foil with circular area of 0.196 cm$^2$ was also investigated (Figure 5B, red curve), showing greater HER activity with a near zero overpotential. The HER polarization curves of the current density are plotted as a function of potential for a representative WS$_2$ sample and shown in Figure 6B (black curve). We find a current density of 10 mA/cm$^2$ at an overpotential of - 350 mV for the WS$_2$ film, comparable to the values observed elsewhere.[40-43] The onset potential obtained for WS$_2$ sample was of −97 mV (vs RHE). Superior catalyst materials give the highest currents at the smallest overpotential. This shows that $WS_2$ films produced through mechanical abrasion are suitable for HER catalyst applications. The figure 6B shows the polarization curves obtained from gold film (substrate used to deposit $WS_2$ samples). A noticeable difference was observed when compared to the gold substrate with the $WS_2$ catalyst, this indicates that the enhanced catalytic performance is from the TMDC. The overpotential is plotted in inset of Figure 6B and the absolute value of the current density within a cathodic potential window with the corresponding Tafel fit is shown, red curve. Thus, the polarization curve shows exponential behaviour, with the Tafel equation *overpotential= a+b log|j|* (where b represents the Tafel slope) and j is the current density. For our $WS_2$ films we find a Tafel slope of 148 mV/dec, see inset of Figure 6B. The reported Tafel slopes vary significantly for different studies depending strongly on the synthesis route of the $WS_2$ films. For example, Bonde et al. reported the HER activity on carbon supported $WS_2$ nanoparticles with Tafel slopes of 135 mV/dec.[41] Xiao et al. used an electrochemical route to obtain the amorphous tungsten sulphide thin films onto nonporous gold, which the Tafel slope was 74 mV/dec.[42] Chen et al. found a similar value (78 mV/dec) for the $WS_2$ prepared at 1000 °C.[43] However, the synthesis route of these works involves high temperature processes and/or several steps to obtain the $WS_2$ catalysts. While the $WS_2$ catalysts exfoliated by mechanical abrasion are rapidly produced through a single low-cost step from cheap and widely available TMDC powders which are already produced on a mass-scale.

**2.4 Triboelectric Nano generator (TENG)**

The triboelectric effect in 2D materials has recently been reported.[44, 45] Previously investigated devices were mainly based on thin- films produced through liquid- phase exfoliation. Here we demonstrate the use of mechanically abraded thin films for the use as TENG electrodes. Figure 7A shows the operation of a simple setup with a thin PET substrate and an abraded nanocrystal film or multilayer stack of abraded 2D materials.

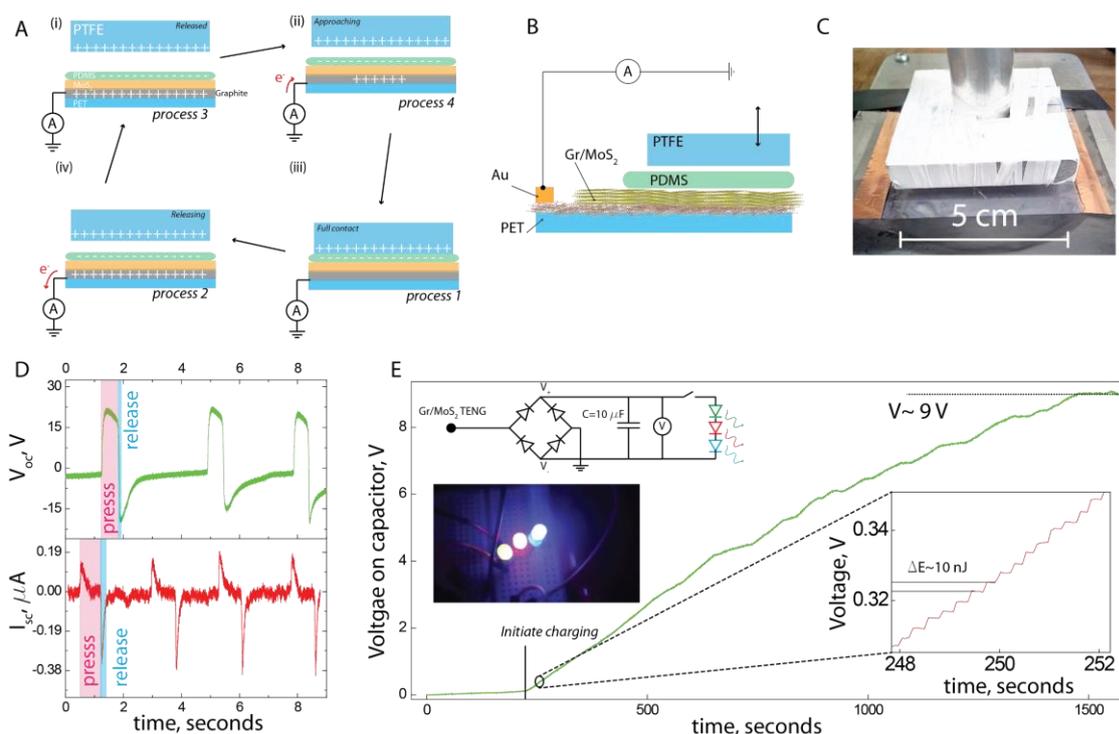

**Figure 7. TENG films based on abraded van der Waals powders.** (A) Schematic showing the evolution of charge within the device. (B) Schematic of the device and electrical setup for acquiring short circuit current. (C) Image of the setup with a Teflon coated aluminium hammer separated from a 1mm thick PDMS film placed on the Graphite-MoS2 heterostructure. (D) Top: temporal response of the open circuit voltage. Bottom: temporal response of the short circuit current. (E) Voltage accumulation on a capacitor vs time. Inset top: rectifying circuit used to charge the capacitor. Inset middle left: three glowing LED's during discharge of the capacitor. Inset right: zoomed in region of the charging curve highlighting the energy stored on the capacitor per cycle.

Typically, high quality TENG devices rely on two materials on the opposite end of the triboelectric series. Recently, it has been demonstrated that a strategy for enhancing the power output relies on the use of multi-layered structures. In this case by introducing charge trapping layers such as MoS2, more charges would be trapped inside the layers via the abundant trap levels that these materials possess [44,45].

In our case we use a PTFE (Teflon) coated hammer, and a fluorinated PDMS pad and an abraded graphite film coated with n-type MoS$_2$. The operation of the device can be explained as follows: after several contacts between both layers, the Teflon pad is completely released from the PDMS pad, which is in turn attached to the MoS$_2$-graphite layer, at this point all layers are neutrally charged, Figure 7A(i); Upon approaching to the PTFE to the PDMS electrons are drawn into the graphitic electrode that is neutralizing, resulting in a positive current Figure 7A(ii). Full contact between these two materials results in charge transfer from one to the other based on the triboelectric series, Figure 7A(iii). Upon releasing, graphitic electrode is electrostatically induced by negatively electrified PDMS, and at this moment, free electrons in it move from electrode to ground, resulting in a negative current Figure 7A(iv).[46-49] Figure 7D shows the open circuit voltage and short circuit current results for three cycles of our proof of concept device which yields voltages in excess of 15 V and short circuit currents of 0.38 $\mu$A, giving a peak power output of 5.7 $\mu$W. To demonstrate the role of MoS2 in improving the output performance of the TENG we compared this device with a TENG device without a MoS2 layer (See supporting information), our initial proof of principle devices shows a ~50% improvement in performance compared to bare graphitic electrodes. Such devices allowed us to use the generated voltage pulses to partially charge a 10 $\mu$F capacitor to 9 V, Figure 7E. The inset shows the energy stored on the capacitor per cycle ($E=1/2C[[V_2]^2 - [V_1]^2]$) when connected via a rectifying diode bridge. Given the wide variety of different two-dimensional materials that can be combined we expect that the efficiency could be significantly improved, thus making abraded 2D material electrodes potential candidates for future flexible energy harvesting TENG electrodes.

## 3. Conclusion

In conclusion we show that a wide variety of functional heterostructure devices can be built up from 2D nanocrystals through a simple mechanical abrasion method. This technology allows for the rapid up-scaling of heterostructures. and we demonstrate its practical use in several simple device applications including conductive few layer graphene coatings and photodetectors. We have extended the technology and demonstrated the successful creation of various more complex vertical heterostructure devices including multi-layer photovoltaics and have shown that abraded WS$_2$ coatings can be used directly as electrocatalysts for the HER reaction as well as TENG electrodes. The ease with which the films can be applied, wide choice of materials and simplicity of up-scalability makes this production route significantly attractive for a large variety of practical applications.

## 4. Methods

### 4.1 Materials

$MoS_2$ (234842-100G), $MoSe_2$(778087-5G) and graphite(282863-25G) powders were purchased from Sigma-Aldrich. $WS_2$ powder was acquired from Manchester nanomaterials and the hBN powder was purchased from Momentive (AC6111). CVD graphene on copper foil was purchased from graphene supermarket. We used specialised tape (Nitto Denko Corporation) ELP-150E-CM for thinning the abraded films and used both commercial PDMS pads PF-30-X4 (retention level 4) and mixed up in house (SYLGARD 184).

### 4.2 Devices fabrication

Devices based on mechanical abrasion are fabricated as described in the main text. For devices including CVD graphene the fabrication was carried out as follows: 950k 6% in Anisole PMMA was spin coated onto CVD graphene on copper, a tape window was then attached and the copper etched away in 0.1 M aqueous solution of ammonium persulfate, which nominally took ∼6 hrs, the CVD graphene was then transferred onto the target device completing the heterostructure. The device along with CVD graphene/PMMA was heated on a hot plate for 1hr at 150 degrees to improve the mechanical contact of the CVD graphene with the abraded nano-crystal films.

### 4.3 Materials characterisation

Raman spectroscopy was carried out using 532 nm excitation at 1 mW laser power which is focused onto a 1 mm spot. AFM was performed using a Bruker Innova system operating in the tapping mode to ensure minimal damage to the sample's surface. The tips used were Nanosensors PPP-NCHR, which have a radius of curvature smaller than 10 nm and operate in a nominal frequency of 330 kHz.

**4.4 Optical measurements**

Optical transmission spectra were recorded using an Andor Shamrock 500i spectrograph with 300l/mm grating and iDus 420 CCD. A fiber coupled halogen white light source was used to excite the photo-active samples which generates 1.4 W at the fibre tip (Thorlabs: OSL2). The white light is collimated to give uniform excitation of 70-100 mW/cm$^2$. The white light source was blocked for the time response using a mechanical shutter with a response time of 10 ms. The spectral dependence of the photocurrent was carried out using 10 nm band pass filters to filter the halogen white light source with the power at each wavelength recorded using a Thorlabs photodiode S120C.

**4.5 Electrical measurements**

Electron transport measurements were carried out using a KE2400 source-meter for both source and gate electrodes. An Agilent 34410A multi meter was used to record the voltage drop over a variable resistor in order to determine the drain current and photo response for different load resistances. Capacitance spectroscopy was performed using a Rhode and Schwarz, Hameg HM8118 LCR Bridge.

Electrochemical data was obtained using an Ivium-stat potentiostat/galvanostat. Linear sweep voltammetry (LSV) experiments were carryout in 0.5 M $H_2SO_4$ with a scan rate of 2 mV/s. For determination of activity of hydrogen evolution reaction (HER), a three-electrode electrochemical cell was used, i.e., saturated calomel electrode (SCE) (reference), platinum foil electrode (counter), and WS2/Au (working). The work electrode area used was 0.147 cm$^2$. The reference electrode was stored in KCl solution and rinsed with deionized water before use. For the measurements, high-purity $N_2$ gas was bubbled into the solution for at least 60 min before the electrochemical measurements. The potentials reported here are with respect to reversible hydrogen electrode (E (RHE) = E (SCE) + 0.273 V.[42])

**Acknowledgements**

F.W acknowledges support from the Royal Academy of Engineering and Royal Society grant RG170424. F.W. and A. De. S acknowledge support from the EPSRC sub-grant EP/S017682/1 Capital Award emphasizing support for early Career researchers. J.F.F acknowledges the Brazilian agencies CNPq (grant number: 430470/2018-5), FAPDF (grant number: 00193-00002066/2018-15 and 193.001.757/2017), for financial support and the research scholarship. SR acknowledges financial support from the Leverhulme Trust (Research grants "Quantum Revolution" and "Quantum Drums"). MF acknowledge financial support from the UK Engineering and Physical Sciences Research Council (EPSRC) (Grant Nos. EP/K017160/1, EP/K010050/1, EP/M001024/1, and EP/M002438/1

**Authors Contribution**

D.N. developed the transfer process for the graphitic electrodes, (together with N. C designed the custom bending rig), carried out device fabrication, measurement, analysis as well as performing atomic force & scanning electron microscopy and Raman characterisation and contributed to writing the manuscript. J.F.F carried out the electrochemical measurements, analysis and contributed to the writing of the manuscript. I. L. and S.R. prepared the lithium perchlorate electrolyte and contributed to writing the manuscript. M.F.C suggested to use the films for TENG electrodes assisted with interpretation of the TENG properties and contributed to writing of the manuscript. S. D.-W developed the TENG measurement setup, interpreted the TENG data and contributed to writing the manuscript. A. De. S. Provided high resolution spectral dependant photocurrent measurements and contributed to writing the manuscript. H.A.F. carried out the initial optical transmission spectroscopy. H. C, N. C and A. W provided technical support. F. W. initiated and supervised the project, contributed to sample fabrication, electron transport, optical measurements, analysis and writing the manuscript.

**Competing financial interests**

The authors declare no competing financial interests.

**Supplementary Information**

More detailed analysis of the optoelectronic performance of the TMDC photodetector and photovoltaics. Resistance vs transparency for graphene conductor. Detailed Raman and AFM study of the different films. Large area method of applying the films and compatibility of $WS_2$ films for the HER reaction with commercially available stainless steel.

# Supplementary Materials: Scalable heterostructures produced through mechanical abrasion of Van der Waals powders

Darren Nutting[1], Jorlandio F. Felix[2], Shin Dong-Wook[3], Adolfo de Sanctis[1], Monica F Craciun[1], Saverio Russo[1], Hong Chang[1], Nick Cole[1], Adam Woodgate[1], Ioannis Leontis[1], Henry A Fernández[1] and Freddie Withers[1*]

[1] *Centre for Graphene Science, College of Engineering, Mathematics and Physical Sciences, University of Exeter, Exeter EX4 4QF, United Kingdom*

[2]*Universidade de Brasília – UNB, Instituto de Física, Núcleo de Física Aplicada, 70910-900 Brasília, DF, Brazil*

[3]*Electrical Engineering Division, Department of Engineering, University of Cambridge, 9 JJ Thomson Avenue, Cambridge, CB3 0FA, UK*

E-mail: f.withers2@exeter.ac.uk

## 1. Example device fabrication

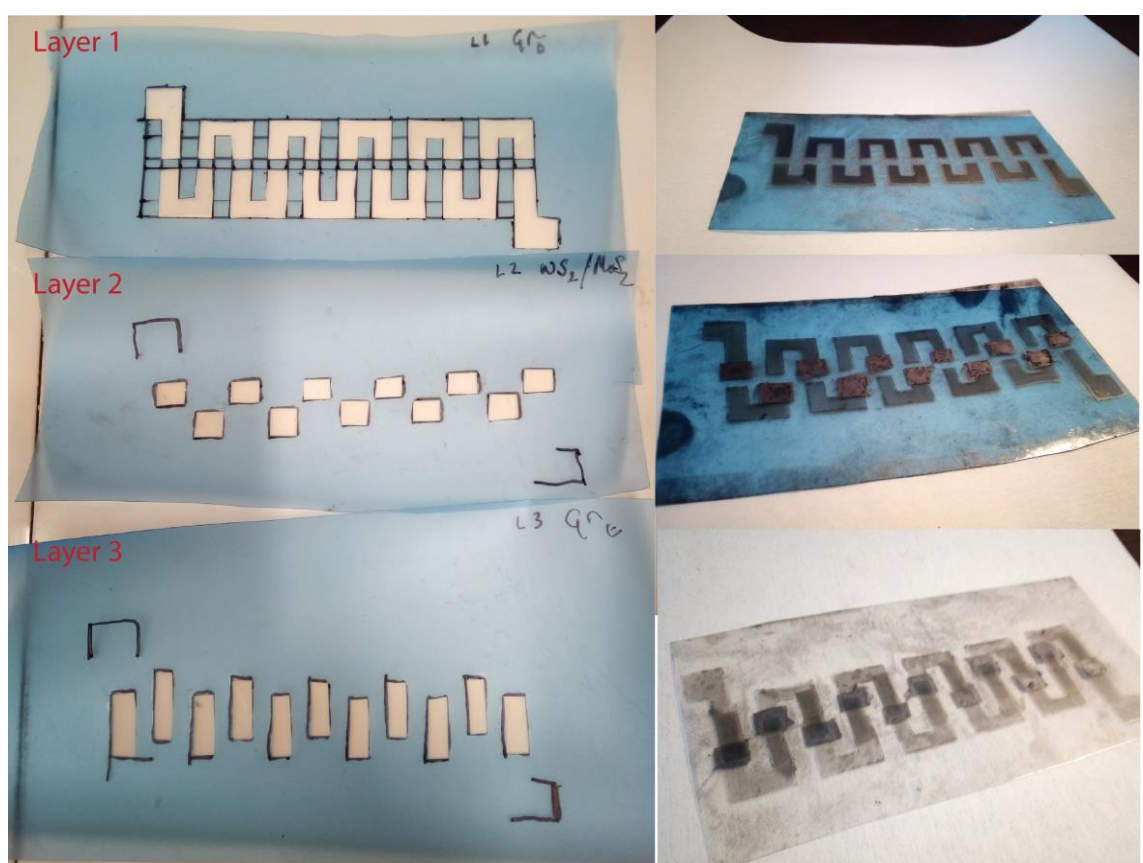

**Figure S1. Example of the multilayer heterostructure production via a tape template.**

The general fabrication process for the multilayer heterostructures is shown in Figure S1. It is currently carried out by cutting templates into the tape and abrading the 2D crystals onto the PET in selected locations which is then repeated with different materials until the heterostructure device is completed.

## 2. Additional optoelectronic devices and measurements

The response for additional photodetector devices are shown in Figure S2A and C.

We also measured a vertical heterostructure device with an amorphous carbon (a-C) top electrode deposited through candle flame deposition[50]. Comparable rise and fall times can be seen for both the simple planar device and the vertical heterostructure with an amorphous carbon (a-C) top electrode.

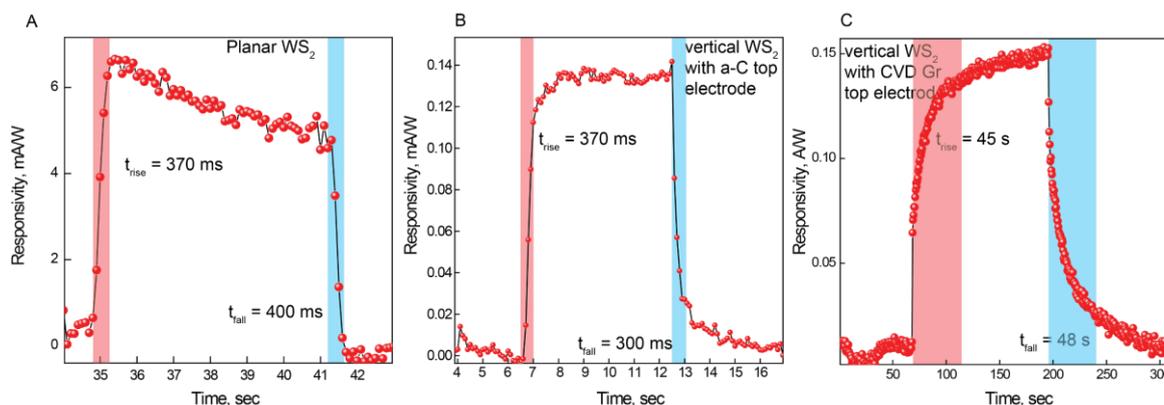

**Figure S2. Response times of different photodetectors.** (A) Planar WS$_2$ junction at V$_{sd}$=50mV. (B) Vertical WS$_2$ junction with a-C top electrode measured at V$_{sd}$=1 V. (C) Vertical WS$_2$ diode with CVD graphene top electrode and Au bottom electrode measured at V$_{sd}$=-1V.

We also produced photodetectors using p and n-type silicon. To produce the devices, we etched away a 1cm$^2$ region of SiO$_2$ in an ammonium flouride:HF (1:1) etch solution. This results in a clean p-/n-Si surface. The WS$_2$ was then abraded directly over the silicon until no pinholes could be observed optically under a 50x objective. CVD graphene was then transferred over the WS$_2$ and baked for 30 min at 100 degrees so the PMMA adheres well to the WS$_2$. For measurements the PMMA was left on, which may have helped bridging any sub micron pinholes in in the WS$_2$ film. The Resistance of the CVD graphene is shown in the inset of Figure S3A and corresponds to R= 650 Ohm. Typical I-V curves in dark and under white light illumination are shown in Figure S3A with the short circuit current and open circuit voltage shown in Figure S3B.

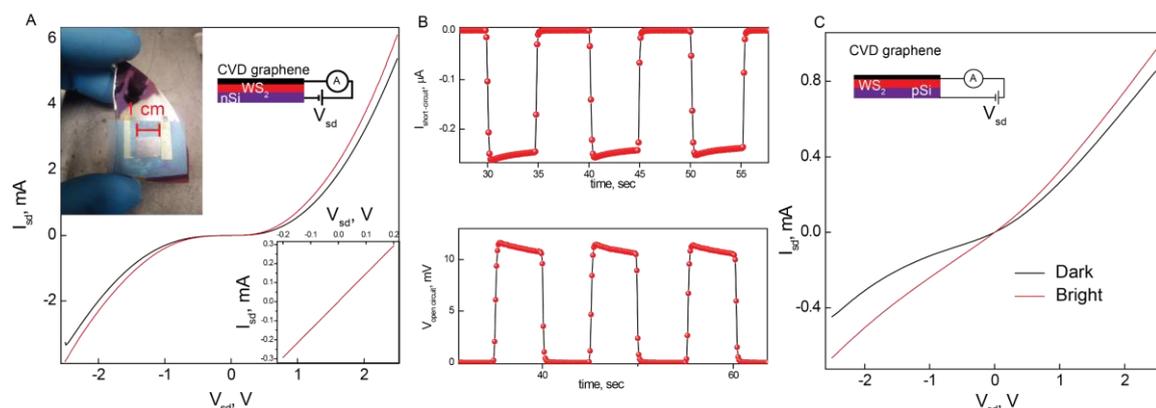

**Figure S3. Photovoltaics with silicon.** (A) Vertical I-V curves for a n-Si / WS$_2$ / CVD graphene device. Inset: (bottom right) I-V of the CVD graphene top electrode R=650 Ohm (top left) image of the device (top right) measurement circuit. (B) Open circuit voltage and short circuit current for the device shown in (A). (C) Vertical WS$_2$ junction but this time with p-type boron doped silicon.

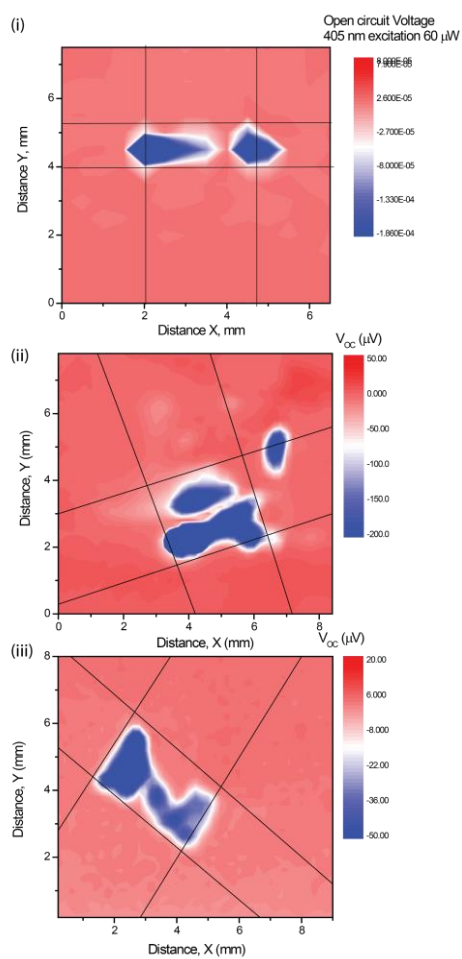

**Figure S4. Photo-voltage mapping for additional devices. (iii) corresponds to a device annealed to 150 degrees for 1hr in ambient conditions exhibiting a much-improved uniformity of the photovoltage.**

Slightly improved performance was observed for p-type silicon. However, overall these devices didn't work well as the TMDC films are rather thick, ~1 $\mu$m. This significantly attenuates light reaching the depletion region. Therefore, most charge separation likely takes place across the Graphene-WS$_2$ Schottky junction.

Figure S4 shows three additional photovoltage maps for devices fabricated through fabrication process 2 as described in the main text of the paper. The devices all show predominant photovoltage over the overlap region between the bottom graphitic electrode, the TMDC barrier layers and the top graphitic electrode. The device shown in (iii) has been annealed to 150 degrees in atmosphere for 60 min and shows more uniform photovoltage over the device area due to better contact quality of the top graphitic electrode. Figure S5 shows additional high spectral resolution photo-responsivity data for the $MoS_2$ planar junction described in Figure 4D of the main text.

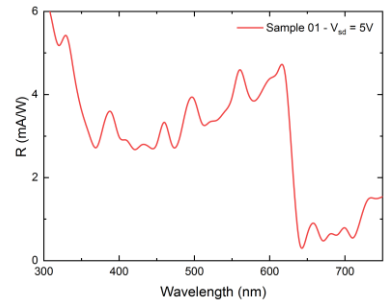

**Figure S5. Spectral dependence of the responsivity for the $MoS_2$ planar detector described in Figure 4D**

## 3. Raman spectroscopy and atomic force microscopy

Raman spectroscopy is an important tool for the characterisation of 2D materials. Figure S6A shows the Raman spectrum of Graphite powder and thin graphitic films on a PET substrate. The normal Raman modes are observed in both cases. The 2D peak at 2711 cm$^{-1}$, G mode at 1581 cm$^{-1}$ and D mode which is associated with disorder at 1350 cm$^{-1}$. The ration of the intensity of the G-mode ($I_G$) to the D-peak ($I_G$) gives a measure of the disorder in our graphitic films. It should be noted that the split peak seen in the G -mode is an artefact of the removal of the PET background which has very closely positioned peaks to the graphite signal and not the D' peak commonly seen at that energy.

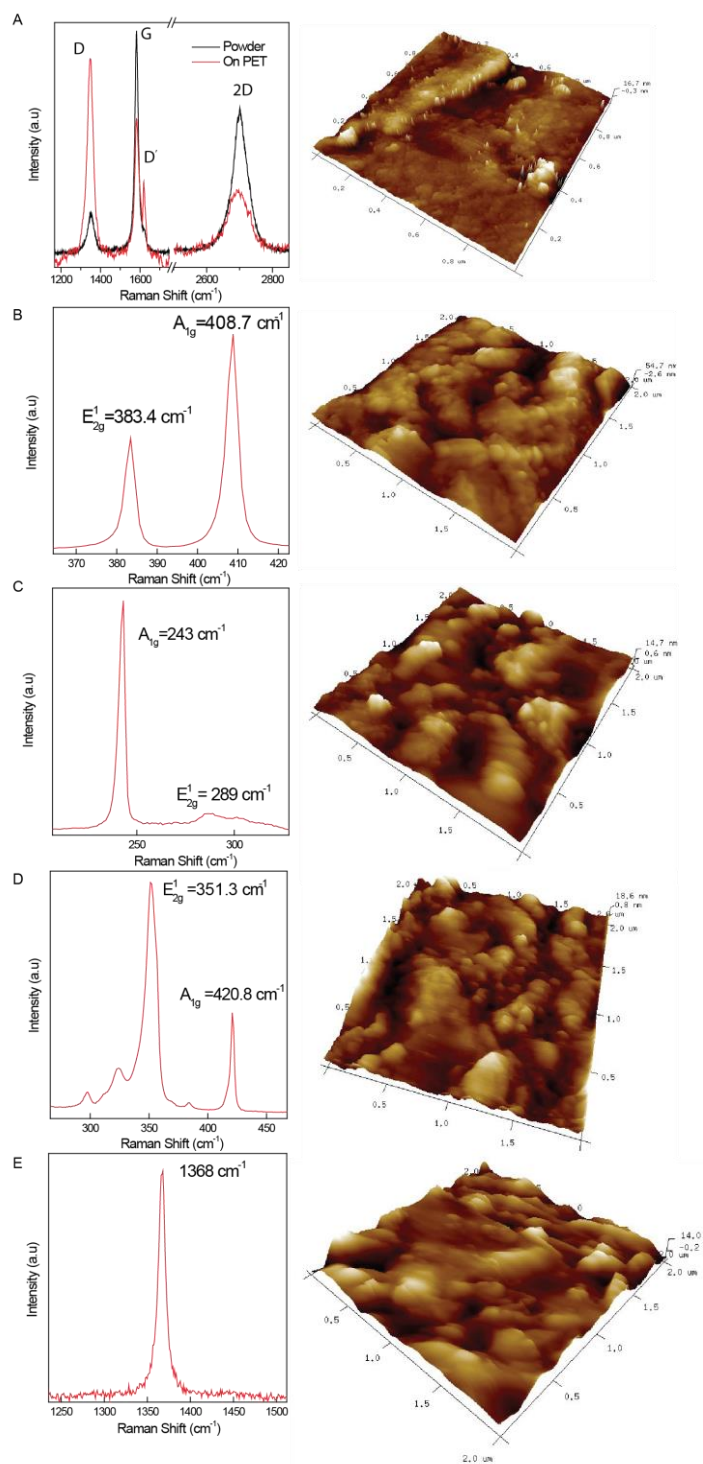

**Figure S6. Raman spectra and AFM.** (A-E) Raman spectra and tapping mode AFM scans for Graphite on PET, MoS$_2$, MoSe$_2$, WS$_2$ and hBN respectively.

In our graphitic powder samples, we obtain an increase of the $I_D/I_G$=0.20, with an increase of the disorder (reduction of crystallite size) for the abraded film $I_D/I_G$=1.43. Refs [9-

[12] of the main text. The Raman spectra for the TMDC films and hBN are all found to be comparable to the literature values for bulk crystals. Estimates of the RMS roughness of each material used are shown in Table T1, and were calculated using Gwyddion analysis software.[51]

| Material | Roughness (nm) | Error (nm) |
|---|---|---|
| $MoS_2$ | 26 | 5 |
| $MoSe_2$ | 4 | 2 |
| $WS_2$ | 6 | 2 |
| hBN | 9 | 3 |
| Graphite | 22 | 10 |

**Table T1. Estimates of surface roughness of various abraded films**

### 4. SEM images of the films

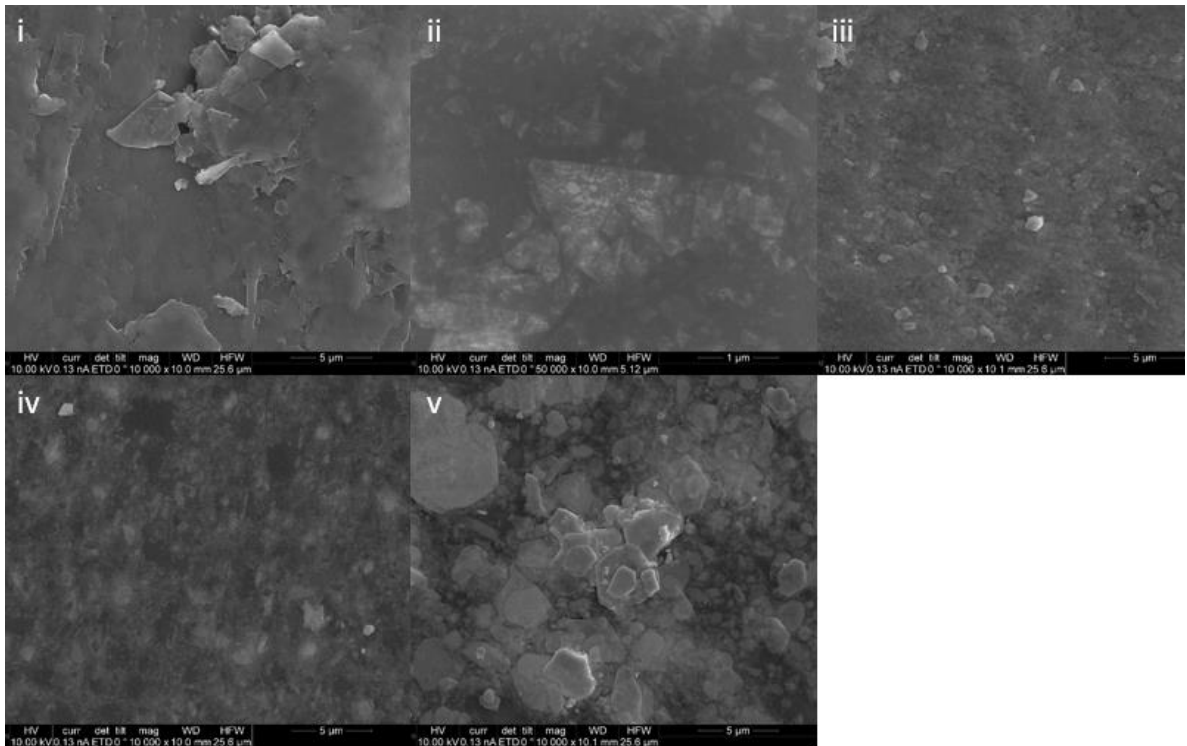

**Figure S7. SEM images.** SEM images of (i) graphite, (ii) hBN, (iii) $MoS_2$, (iv) $MoSe_2$ and (v) $WS_2$ films abraded on a Si/$SiO_2$ substrate.

## 5. Resistance of graphitic films vs transparency

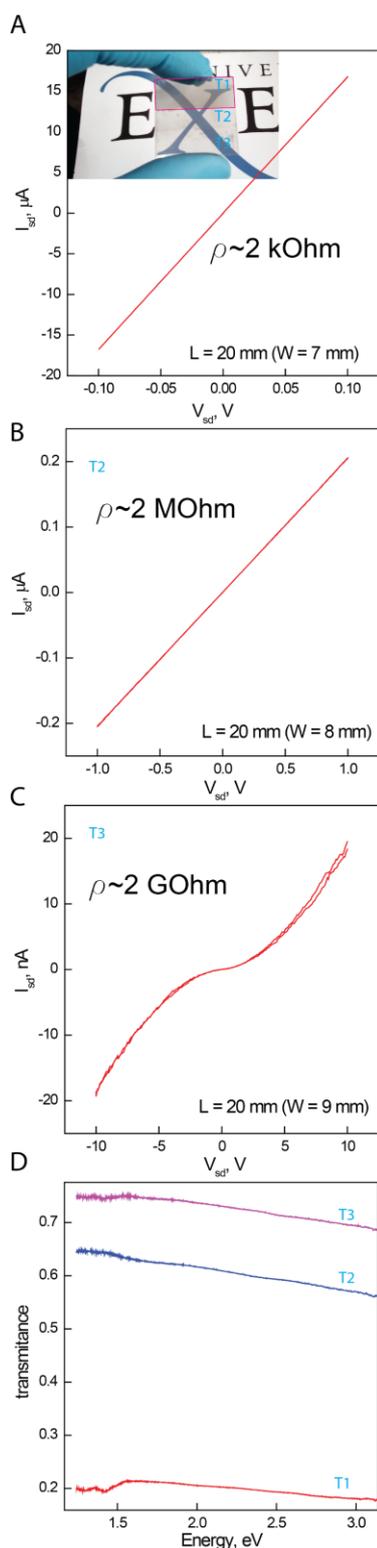

The I-V curves for three different transparency graphitic films are shown in Figure S8(A-C). The transparency is increased by using tape to thin the thickness of the graphitic films. We find a strong dependence of the film thickness with resistivity, increasing by 3-orders of magnitude for each additional peel away of the graphitic film.

The optical transmittance of the films measured in Figure S8(A-C) is also presented in Figure S8D.

**Figure S8. Graphite resistance vs transparency.** (A-C) I-V curves for graphitic films of different thickness. (D) Transparency of the films measured in (A-C).

# 6. Large area films on stainless steel for the hydrogen evolution reaction

For large area films we made with the use of a multi-tool instrument specifically a Titan TTP448HT with oscillation speed of 15000 vibrations per min. Essentially, a PDMS pad is placed on the tool and the 2D powder is placed between the tool and the substrate. Figure S9A shows the general steps for the deposition of a $WS_2$ film on a piece of stainless-steel angle bar. To demonstrate its general use for the HER reaction we sealed a region with the $WS_2$ film with a silicone sealant, which was subsequently filled with DI water (i.e. a makeshift gutter). The photocurrent through a 1MOhm resistor was then recorded and is displayed in Figure S9C for 4 cycles of the incident white light illumination. We find that this method is better suited for large area deposition and could easily be up-scaled easily to the m$^2$.

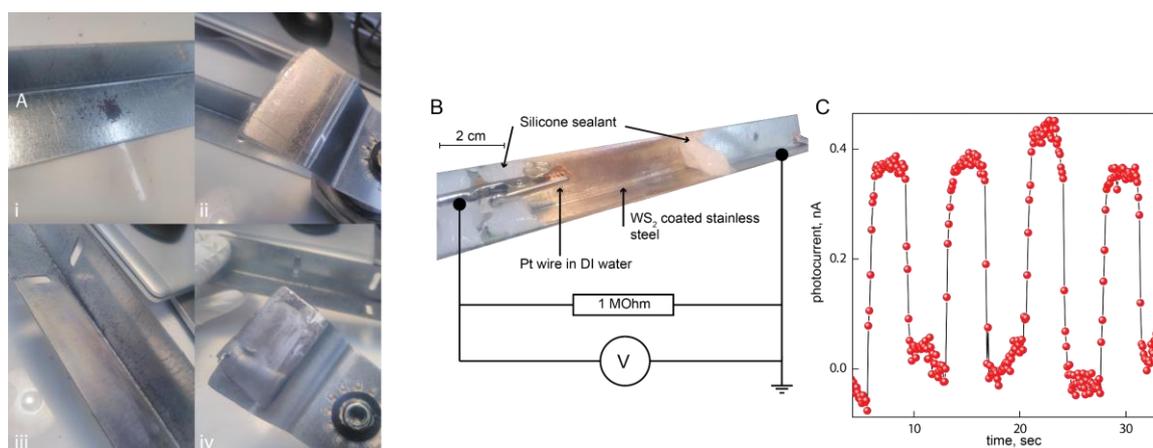

**Figure S9. HER on stainless steel.** (A) Images of the large-scale deposition method. (B) Schematic of the electrical setup and device under illumination. (C) The photocurrent measured through a 1 MOhm resistor when illuminated with a white light source of ~10 mW/cm$^2$.

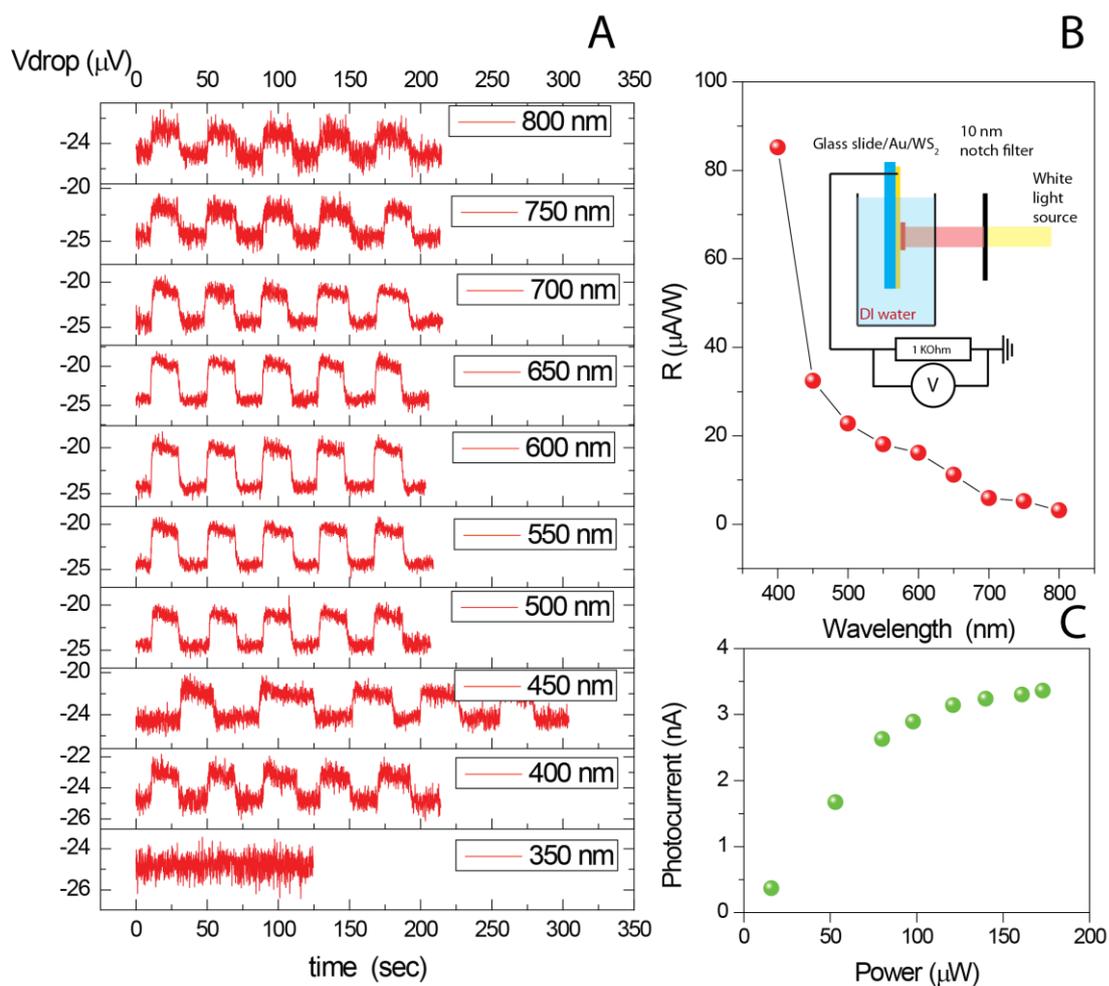

**Figure S10. WS$_2$ catalyst on Au substrate.** (A) temporal response of the photocurrent for a WS$_2$ catalyst film abraded on an Au substrate. (350nm is zero because of the strong absorption in the glass beaker at this wavelength range) (B) Spectral dependence of the responsivity for a WS$_2$ catalyst abraded on Au in DI water. (C) power dependence of the photocurrent at 550 nm excitation.

## 7. Bending rig setup

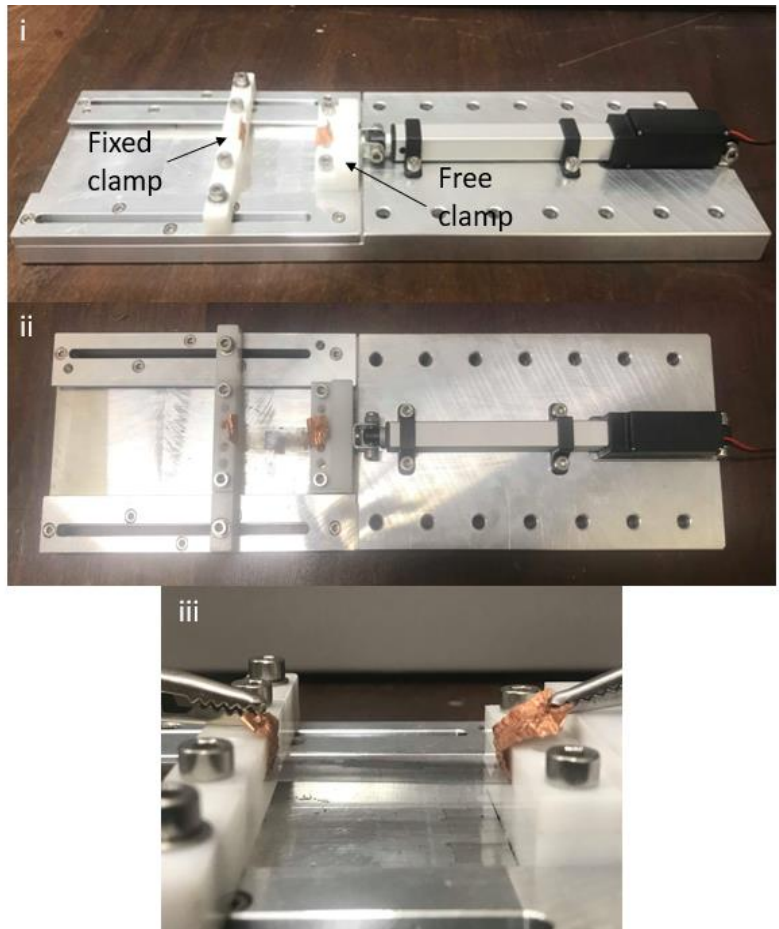

Figure S11 shows the bending rig set up used to apply strain. The devices are held in place by two clamps, one of which is fixed in position to the underlying plate. An Actuonix L16 Micro Linear Actuator is used to push the free clamp forwards in order to bend the sample and apply strain. Dovetail grooves are used to ensure the free clamp can only move parallel to the direction of the actuator arm. The actuator is driven by a voltage pulse and the distance moved by the clamp can then be used to calibrate the rig. The distance moved and therefore strain applied can then by accurately controlled by tailoring the voltage pulse driven through the actuator. Conducting copper tape is used to provide electrical contact to the devices for measurement.

**Figure S11. Image showing the rig used to apply strain to the devices.** (i) and (ii) show the bending rig before and after loading a device. (iii) A zoom in of an unstrained and contacted device within the bending rig.

## 8. Abraded heterostructures as a strain sensor

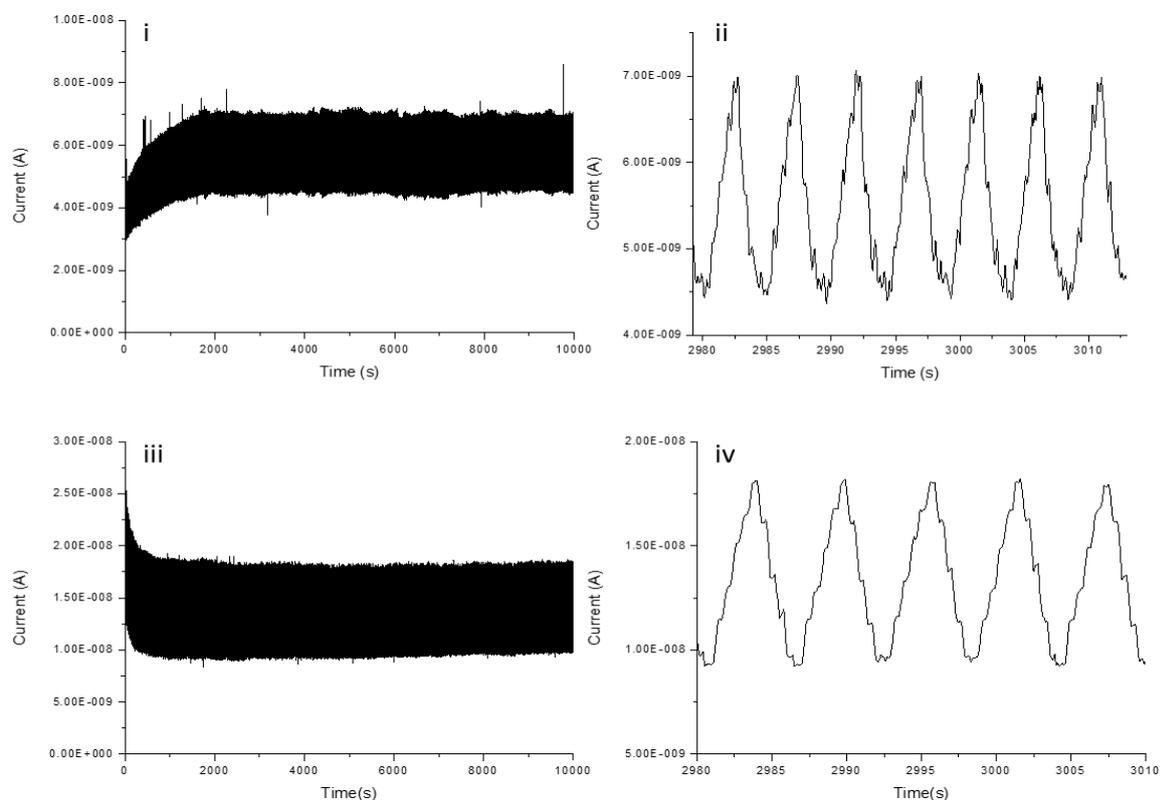

**Figure S12. Strain dependant current and resilience of a Graphite/MoS$_2$/WS$_2$/Graphite heterostructure on a PET substrate.** (i) Change in current with tensile strain at 5 V. (ii) A magnified section of (i) highlighting the individual straining cycles. (iii) Change in current with compressive strain at 5 V. (iv) A magnified section of (iii) highlighting the individual strain cycles.

The devices underwent >1000 strain cycles, with the current quickly reaching a steady state under both tensile and compressive strain. Each cycle involved 10 steps within the strain cycle, resulting in the repeated bumps in figure S12 (ii) and (iv).

## 9. Gr vs Gr/MoS₂ TENG electrodes

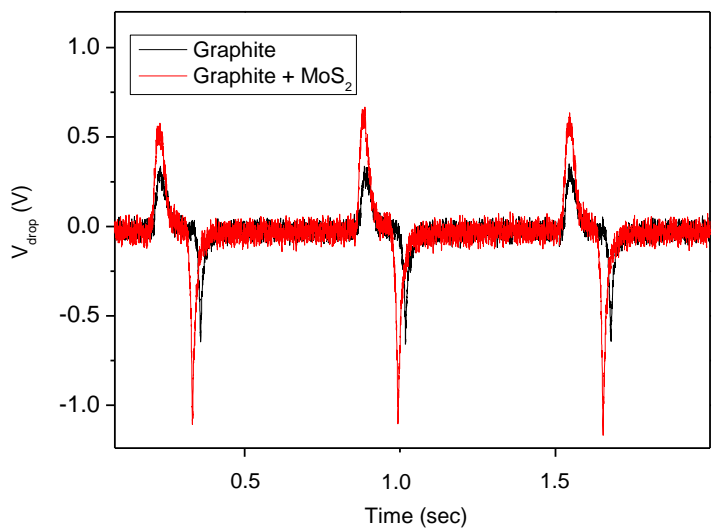

**Figure S13. Voltage drop recorded over a 1MOhm resistor recorded for two TENG electrodes. Graphite (black curve) and Graphite+MoS₂ heterostructure (red curve).**

Figure S13 shows the triboelectric response comparing simple abraded graphite electrodes with graphite + MoS$_2$ heterostructures. We found consistently an improvement in excess of 50% for the Gr/MoS$_2$ heterostructure compared to graphite. This could possible be due to additional electrification at the PDMS/MoS$_2$ interface.

## 9. Electrostatic gating of CVD graphene with abraded hBN dielectrics

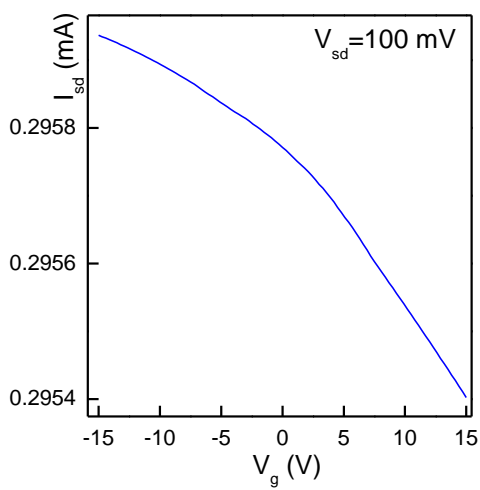

**Figure S14. Electrostatic gating through abraded hBN films. Here V$_g$ is applied to the Au back gate electrode to modulate the source-drain current through the CVD graphene channel.**

Figure S14 shows the weak gate dependence of the resistance of CVD graphene top electrode for the device shown in Figure 2C of the main text. The weak dependence is likely due to the large variance of film thickness and large dielectric thickness of ~ 5 $\mu$m.